\documentclass[slac_one]{revtex4}
\usepackage{graphicx}
\usepackage{fancyhdr}
\begin{document}

\title{Measurements of Single-Top Quark Production at the Tevatron} %% Paper title goes here


\author{Thomas~R.~Junk (on behalf of the CDF and D0 Collaborations)}
\affiliation{FNAL, Batavia, IL 60510, USA}

\begin{abstract}
The latest measurements of the cross sections for singly-produced top
quarks at the Fermilab Tevatron are presented.  The small expected signals
and the large, uncertain backgrounds require careful event selection,
sophisticated methods for separating signal events from background events,
and data-based background predictions and fits.  Both the CDF and D0 collaborations
have solid evidence of single top production, using 2.7 fb$^{-1}$ and
0.9 fb$^{-1}$ of data, respectively.  Both collaborations perform several
analyses to measure the total single-top cross section $\sigma_s+\sigma_t$,
as well as $\sigma_s$ and $\sigma_t$ separately.
\end{abstract}

\maketitle

\thispagestyle{fancy}


\section{INTRODUCTION}

  The observation and precision measurement of the cross sections for single top quark production
have very high priorities at the two Tevatron collaborations, CDF and D0.  Top quarks are expected
to be produced singly via the weak interaction, through $s$-channel or $t$-channel $W$ boson 
exchange, with a total cross section of $\sigma_{st} = 2.86\pm 0.36$~pb~\cite{stxs}, assuming
a top quark mass of 175~GeV.  
The discovery of the top quark~\cite{topdisc} was in $t{\bar{t}}$ production via the strong interaction
since this mode has a more distinct final state (two massive top quarks instead of
just one), and it has a higher cross section.  The strong interaction production processes
are fairly well understood,
and it appears from them that the top quark is the spin-1/2 SU(2) partner of the bottom quark.  
Nonetheless, the top quark's weak interactions are less well constrained by measurements.
A precise measurement of single top quark production can be interpreted directly in terms of
$|V_{tb}|$.  Furthermore, nonstandard
production mechanisms for single top may enhance or suppress the separate cross sections
$\sigma_s$ and $\sigma_t$~\cite{taityuan}.  The simplest such model is one in which a fourth
generation of quarks raises the rank of the CKM matrix from three to four, thus removing the
$3\times 3$ unitarity constraint which provides, along with precise measurements of other
CKM matrix elements, nearly all of our knowledge of $|V_{tb}|$.  If a fourth generation of
quarks exists, $|V_{tb}|$ could be measurably lower than unity~\cite{maltoni}.  On the other
hand, an enhancement in $|V_{ts}|$ is possible if a fourth generation is present, although
this is increasingly well constrained by precision measurements~\cite{pdg}.
Other models including such particles as extra singlet quarks, stop quarks, $W^\prime$ bosons,
and flavor-changing neutral currents mediated by gluons would have profound, measurable effects
on the single-top quark production cross sections.

  Observation and measurement of singly-produced top quarks is also important from a purely
technical standpoint.  Because of the small size of the signal compared with not only the
backgrounds but the systematic uncertainties on the backgrounds, the observation of single top
quarks must be done with sophisticated multivariate techniques.  These techniques have been
in use in HEP for many years but continuously undergo further development, and the single top
searches at CDF and D0 use the latest innovations, described here.  These techniques will also be
needed at the LHC, as the signal-to-background ratios for many interesting physics processes is
very small.  Furthermore, the dominant final state used for searching for single top production
is also an important final state for searching for
a Standard Model Higgs boson with a mass between 114 and 140 GeV.

\section{DATA, MONTE CARLO, and ANALYSIS TECHNIQUES}

The CDF data sample size for the measurements presented here is 
%($2.7\pm 0.16$)~fb$^{-1}$~\cite{CDF},
%and the D0 data sample size is ($0.89\pm 0.05$)~fb$^{-1}$~\cite{d0prd}. 
2.7~fb$^{-1}$~\cite{CDF},
and the D0 data sample size is 0.9~fb$^{-1}$~\cite{d0prd}. 
Both collaborations select events with an isolated, energetic lepton, two or more jets, at least
one of which is b-tagged, and a large missing transverse energy.  The selection requirements
differ somewhat between the collaborations due to differences in the detectors and the choice
of operating points.  CDF requires leptons to have $P_T>20$~GeV, while D0 requires $P_T>15$ GeV.
The $\not\!\! {E}_T$ requirement is also looser for D0: 15~GeV as compared to CDF's 25~GeV.  A feature
of the $t$-channel signal is that the top quark often recoils against a light-flavored jet
which is preferentially located at high pseudorapidity, $|\eta|$.  In order to maximize the sensitivity to
such events, the collaborations extend their jet coverage to $|\eta|<2.8$ (CDF) and 
$|\eta|<3.4$ (D0).  The channels are separated
by the number of jets and the number of b-tags, and the results are combined statistically at the end.

The background levels are quite high, with approximately 80\% coming from $W$+jets final
states in the 2-jet sample, and 47\% from $W$+jets final states in the 3-jet sample, with
most of the remainder coming from $t{\bar{t}}$ events.  Only about half of the selected
$W$+jets background events have $B$ hadrons in them, even though they are $b$-tagged.
In order to provide further separation of the $W+b{\bar{b}}$ and signal events
from $W+c{\bar{c}}$, $W+c$ and $W+$light-flavored-jets events, a neural network flavor separator~\cite{kfs} is
used by CDF as a continuous input variable to all of the multivariate analyses.
The multivariate analyses described below are critical in order to obtain sufficient
sensitivity to test for the presence of single top production.

The single top signal is characterized by several prominent features which can be exploited to
separate it from the background.  The mass of the top quark is already precisely measured, and
signal events contain just one top quark.  Hence, $m_{\ell\nu b}$, where the $\nu$ momentum is
obtained from the missing transverse energy $\not\!\! {E}_T$ and an $m_W$ constraint, is a powerful variable.  
Another variable is $Q\times\eta$,
the charge of the lepton times the pseudorapidity of the untagged jet~\cite{cpyuan}.  This variable
is powerful because $t$-channel single top production involves radiation of a $W$ boson from
a valence quark of the proton or antiproton; the $W$ quark then collides with a sea $b$ quark in the
other beam to make a single top quark.  The recoiling valence quark is usually a light-flavored quark
which has a large momentum component along the beam axis in its original direction of motion.  The
charge of the lepton from $W$ decay from the top quark decay is strongly correlated with whether the
original beam quark radiating the $W$ was in the proton or the antiproton.  These correlations do
not exist for background $W$+jets events or $t{\bar t}$ events, and the jet $\eta$ distributions for
background events are more central.  Furthermore, top quarks produced singly are expected to be produced
nearly 100\% polarized~\cite{mahlonparke}.  Variables taking advantage of the resulting correlations between
the directions of the leptons and the jets in these events are used.  Not every variable needs to key on
a particular feature of the signal prediction.  The variable $m_{jj}$, the invariant mass of the two jets
in $W$+two-jet events, has a more distinct distribution for the backgrounds, peaked at low masses, than
for the signal, which has a broader distribution.

The collaborations use several techniques for extracting the signal from the backgrounds --
boosted decision trees (BDT), matrix elements (ME), neural networks (NN) and likelihoods (LF).  The same events
are analyzed by each discriminant, and the input variables to the discriminants are very similar
from one discriminant to another.  Boosted decision trees have the advantage that the addition of
variables that do not carry useful information does not adversely impact the sensitivity.  D0 uses
Bayesian Neural Networks~\cite{bnn} which are highly optimal and resistant to overtraining,
and CDF uses NeuroBayes program~\cite{neurobayes} which has a sophisticated variable preprocessing
step which also improves sensitivity while reducing the possibility of overtraining.

The matrix element technique involves integrating posterior probability densities computed with leading-order
matrix elements squared for each known process, integrated over possible true values of the
parton momenta, using a transfer function to express the probability density of observing
a particular kinematic configuration after reconstruction, as a function of the kinematic
configuration of the underlying partons.  In this sense it is different from the other techniques
in that the variables input to the boosted decision trees and the neural networks are the best
estimates of the parton momenta, while the matrix element integrates over all possible parton momenta.
CDF also has an analysis using a projective likelihood technique~\cite{CDF} commonly used at LEP.

The different techniques bring sensitivity to different features of the signal and the backgrounds.
In order to obtain the most sensitivity to the signal, combinations of the several analyses' results
are performed within each collaboration to produce combined cross sections and significance levels.

Estimations of the systematic uncertainty are quite mature in the single top analyses.  Because the
background rates are so large and uncertain, the constraints from the data using the histograms
of the discriminant outputs are particularly important.
Sources of uncertainty like the jet energy scale
(which affects $m_{\ell\nu b}$ and $m_{jj}$ quite strongly), initial- and final-state radiation,
PDF's and variable mismodeling introduce effects both on the predicted signal and background acceptances,
and also the shapes of their contributions in the final discriminant histograms.  All sources of uncertainty
are included with their proper correlations.

\begin{figure}[h]
\centering
\includegraphics[width=48mm]{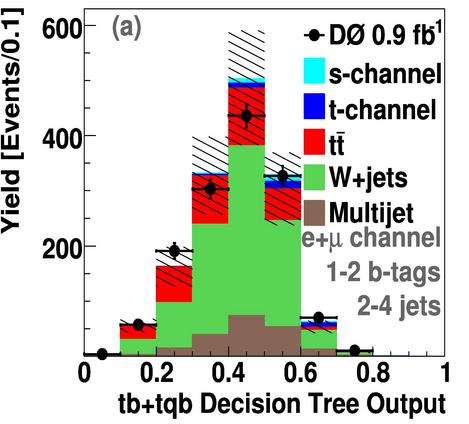}
\includegraphics[width=50mm]{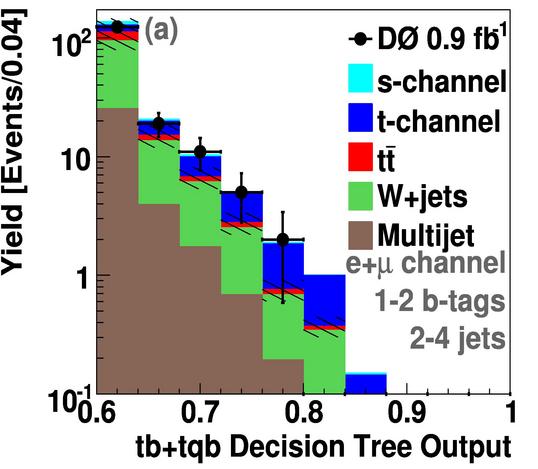}
\includegraphics[width=50mm]{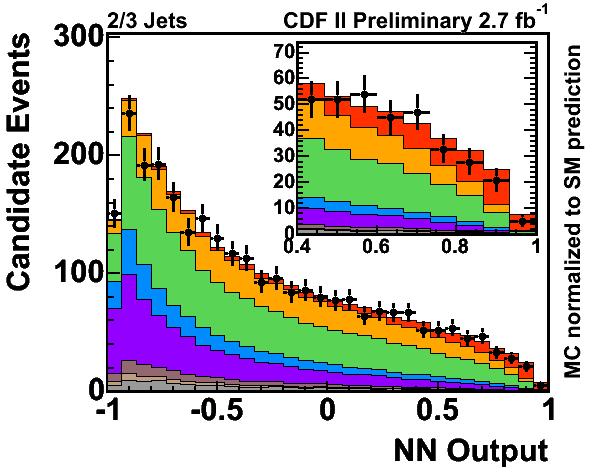}
\includegraphics[width=10mm]{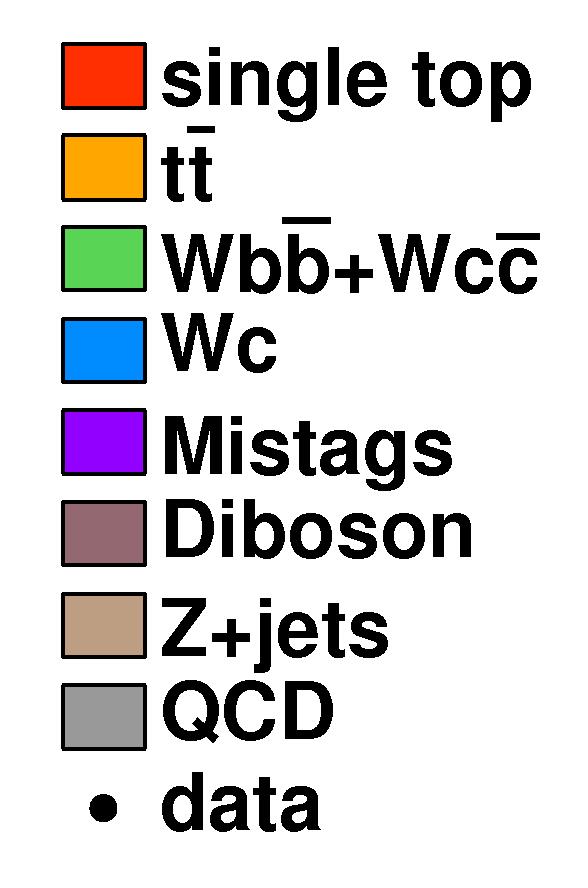}
\caption{Left two graphs: distribution of D0's BDT search results.  Points represent
the data, and the background and signal contributions are shown as colored, stacked histograms.    
The middle graph is a zoom-in of the high-score region. The signal shown is the best fit to the data.
 Rightmost graph:  distribution of CDF's neural network discriminant.  
The signal shown is the SM prediction.} \label{disc}
\end{figure}

Figure~\ref{disc} shows the resulting discriminant distributions of two representative
analyses; the BDT analysis of D0 and the NN analysis of CDF.

\section{CROSS SECTION RESULTS AND SIGNAL SIGNIFICANCE}

\begin{figure}[h]
\centering
\includegraphics[width=47mm]{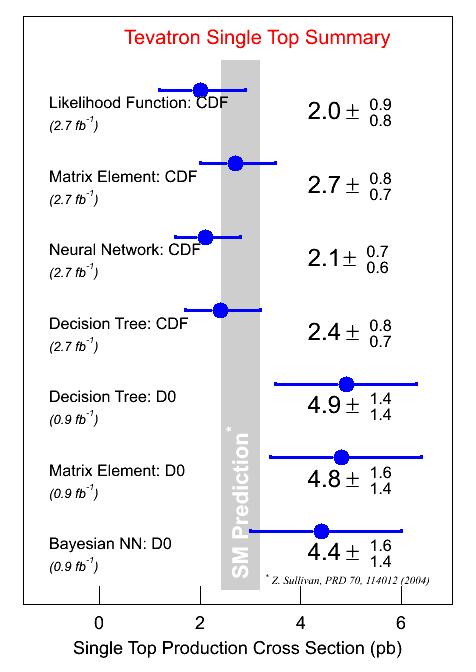}
\begin{minipage}[b]{51mm}
\includegraphics[width=36mm]{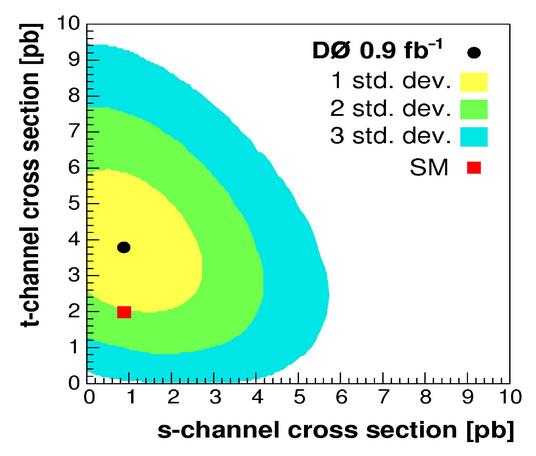} \\
\includegraphics[width=42mm]{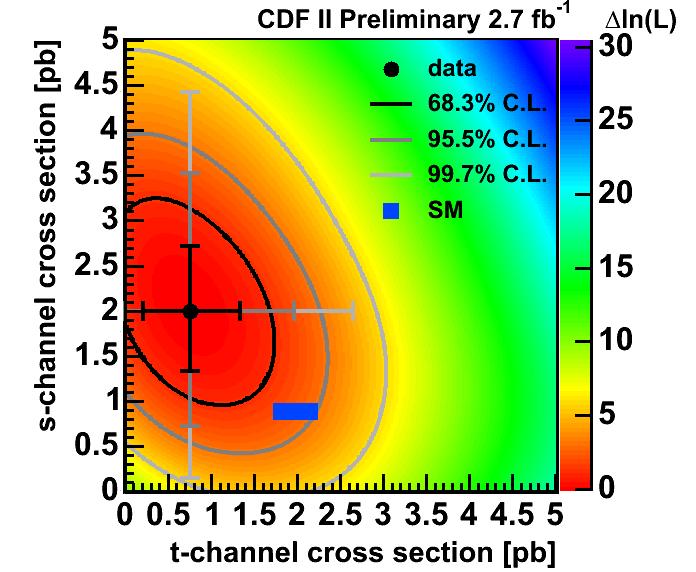}
\end{minipage}
\caption{Left: Summary of single-top cross section measurements by CDF and D0.  The four CDF measurements are
highly correlated with each other, and the three D0 measurements are also highly correlated with each other.
The SM NLO prediction of the cross section is also shown.  Right:  Two-dimensional posterior plots showing
the consistency of the data with single-top production as a function of both $\sigma_s$ and $\sigma_t$.  Top:
D0's BDT analysis, bottom: CDF's NN analysis.  Note that the axis label conventions in the two-dimensional
CDF and D0 plots are opposite.} \label{xssummary}
\end{figure}

Both collaborations use a Bayesian technique to extract cross sections, integrating over systematic
uncertainties~\cite{pdg}, using flat priors in the cross section $\sigma_{st}$.
The measurements for each channel are summarized in Figure~\ref{xssummary}.  Two-dimensional extractions
of $\sigma_s$, the $s$-channel cross section vs. $\sigma_t$, the $t$-channel cross section, are also
obtained by both experiments, separately for each analysis channel.  Two examples are shown in
Figure~\ref{xssummary}, for the BDT analysis of D0, and for the NN analysis of CDF.

D0 combines results using the best linear unbiased estimator technique~\cite{blue}, and obtains
the value $\sigma_{st}=4.7\pm 1.3$~pb.   CDF also combines its measurements in this way,
but CDF in addition forms a super-discriminant out of the discriminants from the
contributing analyses in order to make a single super-analysis which is more sensitive than any of
the contributing ones.  CDF's super-discriminant is built using an evolutionary neural network~\cite{neat}~\cite{CDF},
which is trained to optimize the expected discovery significance, instead of the usual classification
error function.  CDF's combinations were performed for the 2.2 fb$^{-1}$ results, but were not finished
for the 2.7 fb$^{-1}$ results.

The two collaborations take similar approaches to estimate the significance of the excess of data over
the background predictions.  D0 computed the combined cross section in a large sample of background-only
Monte Carlo pseudoexperiments and computed the $p$-value to the the fraction of pseudoexperiments with a measured
cross section greater than or equal to the observed one.  D0 obtains a $p$-value of 0.00014, which corresponds
to a $3.6\sigma$ excess, with a median expected $p$-value of $2.3\sigma$, assuming the SM cross section
for single-top production.  CDF calculates $p$-values using the likelihood ratio test statistic~\cite{pearson},
again using a large sample of background-only pseudoexperiments.  CDF expected to see a $5.0\sigma$ excess in the
median signal-like outcome, and observed a $3.7\sigma$ excess.

\section{SUMMARY}

Significant progress has been made by the CDF and D0 collaborations to observe single top production and to
measure its total cross section, and also the two cross sections $\sigma_s$ and $\sigma_t$ separately.  The goal
is to observe the production at the $5\sigma$ level of significance or greater, and to measure the cross sections
with the highest possible precision.   These measurements help constrain models of new physics affecting the
weak interactions of top quarks, and also models in which exotic particles contribute to the selected single-top
quark sample.


\begin{thebibliography}{99}

\bibitem{stxs} B.~W.~Harris {\it et al.}, Phys. Rev. D {\bf 66}, 054024 (2002); \\
Z.~Sullivan, Phys. ReV. D {\bf 70}, 114012 (2004).

\bibitem{topdisc} F.~Abe {\it et al.} (CDF Collaboration), Phys. Rev. Lett.
{\bf 74}, 2626 (2005); \\
S. Abachi {\it it al.} (D0 Collaboration), Phys. Rev.
Lett. {\bf 74}, 2632 (1995).

\bibitem{taityuan} T.~Tait and C.-P.~Yuan, Phys. Rev. D {\bf 63}, 014018 (2001).

\bibitem{maltoni}  J.~Alwall {\it et al.},
  %``Is V(tb) ~= 1?,''
  Eur.\ Phys.\ J.\  C {\bf 49}, 791 (2007).

\bibitem{pdg} C. Amsler {\it et al.}, Physics Letters B {\bf 667}, 1 (2008).

\bibitem{CDF} CDF Collaboration, conference notes 9479, 9451, 9464, 9445, and 9460 (2008); \\
CDF Collaboration, T. Aaltonen {\it et al.}, ArXiv:0809.2581 [hep-ex] (2008), submitted to Phys. Rev. Lett.

\bibitem{d0prd}    V.~M.~Abazov {\it et al.}  [D0 Collaboration],
  %``Evidence for production of single top quarks and first direct  measurement
  %of |V(tb)|,''
  Phys.\ Rev.\ Lett.\  {\bf 98}, 181802 (2007); \\
V.~M.~Abazov {\it et al.}  [D0 Collaboration],
  %``Evidence for production of single top quarks,''
  Phys.\ Rev.\  D {\bf 78}, 012005 (2008).

\bibitem{kfs} S.~Richter, Ph.D. thesis, University of Karlsruhe (2007).

\bibitem{cpyuan} C.-P.~Yuan, Phys. Rev. D {\bf 41}, 42 (1990).

\bibitem{mahlonparke}   G.~Mahlon and S.~J.~Parke,
  %``Improved spin basis for angular correlation studies in single top quark
  %production at the Tevatron,''
  Phys.\ Rev.\  D {\bf 55}, 7249 (1997).

\bibitem{bnn}  R.~M.~Neal, {\it Bayesian Learning of Neural Networks}
(Springer-Verlag, New York, 1996); \\
P.~C.~Bhat and H.~B.~Prosper, in {\it Statistical problems in Particle Physics,
Astrophysics and Cosmology}, edited by L.~Lyons and M.~K.~\"Unel (Imperial College
Press, London, England, 2006), p. 151.

\bibitem{neurobayes} M.~Feindt and U.~Kerzel, 
Nucl. Instrum. Methods A~{\bf 559}, 190 (2006).

\bibitem{blue}  L.~Lyons, D.~Gibaut and P.~Clifford,
  %``HOW TO COMBINE CORRELATED ESTIMATES OF A SINGLE PHYSICAL QUANTITY,''
  Nucl.\ Instrum.\ Meth.\  A {\bf 270}, 110 (1988).

\bibitem{neat} K.~O.~Stanley and R.~Miikkulainen, Evolutionary Computation {\bf 10 (2)} 99-127 (2002).

\bibitem{pearson} J.~Neyman and E.~Pearson, Philos. Trans. R. Soc. London Ser. A~{\bf 231} 289 (1933).

\end{thebibliography}
\end{document}